\UseRawInputEncoding
\documentclass[aps,pre,showpacs,preprint,superscriptaddress]{revtex4}
\usepackage{graphicx}
\usepackage{subfigure}
\usepackage{amsmath}
\usepackage{ulem}
\usepackage{bm}
\usepackage{amssymb}
\usepackage{color}
\usepackage{epstopdf}
\usepackage{epsfig,dsfont,multirow}
\pdfoutput=1
\begin{document}
\title{Attosecond electron bunch generation by an intense laser propagation in conical channel with a curved wall}
\author{Min Zhang}
\affiliation{Key Laboratory of Beam Technology of the Ministry of Education, and School of Physics and Astronomy, Beijing Normal University, Beijing 100875, China}
\author{Cui-Wen Zhang}
\affiliation{Key Laboratory of Beam Technology of the Ministry of Education, and School of Physics and Astronomy, Beijing Normal University, Beijing 100875, China}
\author{De-Sheng Zhang}
\affiliation{Key Laboratory of Beam Technology of the Ministry of Education, and School of Physics and Astronomy, Beijing Normal University, Beijing 100875, China}
\author{Hai-Bo Sang}
\affiliation{Key Laboratory of Beam Technology of the Ministry of Education, and School of Physics and Astronomy, Beijing Normal University, Beijing 100875, China}
\author{Bai-Song Xie \footnote{corresponding author: bsxie@bnu.edu.cn}}
\affiliation{Key Laboratory of Beam Technology of the Ministry of Education, and School of Physics and Astronomy, Beijing Normal University, Beijing 100875, China}
\affiliation{Institute of Radiation Technology, Beijing Academy of Science and Technology, Beijing 100875, China}
\date{\today}
\begin{abstract}
By using two-dimensional particle-in-cell simulations, attosecond electron bunches with high density, high energy and small divergence angle can be obtained by $p$-polarized laser irradiation in conical channel with curved wall. We find that some electrons in the wall are pulled into the channel by the transverse electric field and are directly accelerated. Meanwhile, they move steadily along the conical wall via laser pondermotive force. The results show that the focusing effect of the curved wall conical channel is stronger than that of the traditional flat wall conical channel, and the density of the attosecond electron bunches is increased by nearly $175\%$ as well as the maximum energy is increased by $36\%$. We also find that the quality of the electron bunches is affected by the geometry of the concial channel wall. Interestingly it is found that the attosecond electron bunches obtained from the specific concial channel with the hyperbolic geometry of the curved wall can keep stable around the maximum electron energy within $10T_0$ even if they have left the channel.

\end{abstract}

\maketitle

\section{Introduction}

High-power lasers make it possible to generate ultrashort electron bunches through laser-plasma interactions \cite{Mourou:2006zz,Maine:1988gyb,Strickland:1985gxr,Danson:2019gxr}. It is well known that the ultrashort electron bunches is of great importance for ultrafast science because they can be used in many ultrafast applications such as electron imaging, electron diffraction and spectroscopy \cite{Dpp:2024zz,Zhao:2018qgm,Bai-Fei:2021lco}.

It is noted that in the interaction by the intense laser with low-density plasma, a single
electron bunch can be generated by laser-wakefield acceleration mechanism\cite{Sprangle:1989nf,Umstadter:1996aa,Luttikhof:2009ucr,Li:2013rux}, while the length of the bunch is typically not shorter than $1-5 \rm{\mu m}$ and lasts only a few femtoseconds.
On the other hand, however, for the intense laser interacting with high-density plasma, attosecond electron bunches can be generated through two mechanisms, the $\textbf{v} \times \textbf{b}$ heating \cite{Kruer:1985bk} and the vacuum heating \cite{Brunel:1987bk,Pukhov:1999zz,Tsakiris:2000zz}. In 2004, Naumova et al \cite{Naumova:2004zz} showed a $p$-polarized intense laser interacting with a sharp boundary of an overdense plasma channel to generate attosecond electron bunch. However, the energy of the attosecond electron bunch is only $20-30\rm{MeV}$. Later, Yu et al \cite{Hu:2015zz} proposed to use an intense laser to hit the traditional flat wall conical channel and obtained spatially periodic attosecond electron bunches which have an average density of about $10 n_c$ and cut-off energy of $380\rm{MeV}$. Obviously the results show that the conical channel can effectively focus the laser and generate attosecond electron bunches with higher density and energy. Besides, the smaller the opening of the conical channel is, the more obvious the focusing effect exhibits, but the feebler the collimation of the electron bunches are. For example, the average emission angle is about 25$^{\circ}$ when the right opening diameter of conical channel is $D_r=2 \rm{\mu m}$. This would be unfavorable for potential applications to get highly collimated attosecond electron bunches.

In this paper, in order to obtain high quality attosecond electron bunch, we design a conical channel target with a wall of hyperbolic curve given by $y=a/x+b$, which is called the curved wall conical channel. For the sake of optimum target configuration, we have made a comparison study about the quality of electrons generated by the interaction of intense laser with the curved wall conical channel and the flat wall conical channel. It is found that the average density and the cut-off energy of the electron bunches in the curved wall case is $20n_c$ and 400MeV, respectively, better than those in the flat wall case. And the electron bunches have a narrower longitudinal thickness, i. e., $\sim 300\rm{as}$. In addition, we also reveal how the geometry of different curved wall influence the energy and density of the attosecond electron bunches. These electron bunches are believed to be helpful as a new tool to make some new researches and applications such as in attosecond electron diffraction, microscopy and ultrashort coherent x-ray generation via Thomson scattering \cite{Wu:2010di,Wu:2011zzk}.

The paper is organized as follows. In Sec. II, the simulation model is introduced and the results are presented with a physical analysis. In Sec. III, the results of other types of the curved wall conical channels are shown comparatively. Finally, a brief summary is given in Sec. IV.

\section{SIMULATION MODEL AND RESULTS}

\begin{figure}[ht!]\centering
\includegraphics[width=0.5\textwidth]{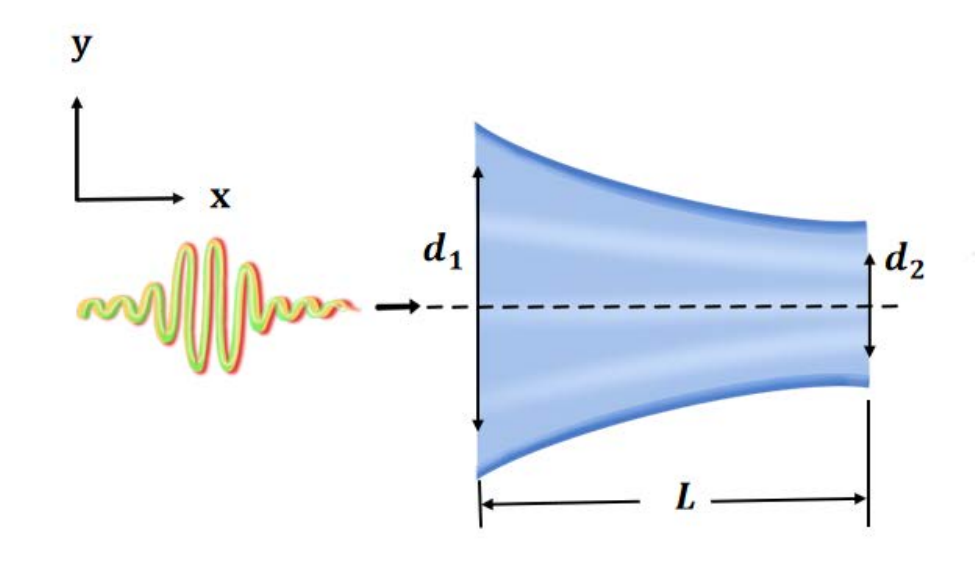}
\caption{\label{FIG1}   The schematic diagram of an attosecond electron bunches generated by the interaction of an ultra-intense laser with a gold curved wall conical channel and the geometry of the curve wall is hyperbola. The parameters of the conical channel are: $d_1=10\rm{\mu m}$, $d_2=5\rm{\mu m}$ and $L=25\rm{\mu m}$. }
\end{figure}

We take two-dimensional particle-in-cell (2D PIC) simulations with open-source code EPOCH \cite{Ridgers:2013kjw,Ridgers:2015ikp}. As indicated in Fig.\ref{FIG1}, a $p$-polarized Gaussian laser impacts a conical channel and the laser propagates along the $x$-axis from the left side of the simulation box, meanwhile the optical axis is consistent with the central axis of the conical channel. The dimensionless laser field is $a=a_0\exp(-(t-t_0)^2/\tau^2) \exp(-y^2/w^2)$, where $a_0=eE_0/m_ec\omega_0=30$ is the laser intensity, $\omega_0=2\pi/T_0$ is the laser frequency, $\tau=4T_0$ is the pulse duration, $w=4\lambda_0$ is the laser focal spot radius and $\lambda_0=cT_0=1\mathrm{\mu m}$ is the laser wavelength. And $e$, $m_e$ and $E_0$, $c$ are the electron charge and mass, electric field amplitude, and the speed of light in the vacuum, respectively.

The curved wall conical channel consists of an initially neutral mixture of electrons and gold ions charged by $+1$ unit and the plasma density is $100n_c$, where $n_c=m_e\omega_0^2/4\pi e^2=1.1\times10^{21}\rm{cm^{-3}}$ is the critical density. The axial length of conical channel is $25\lambda_0$ located from $x=10\lambda_0$ to $x=35\lambda_0$, we guarantee the slope of the curve is zero at $x=35\lambda_0$. The diameter of the left and right opening of the conical channel is $d_1=10\rm{\mu m}$ and $d_2=5\rm{\mu m}$, respectively. The thickness is $0.2\lambda_0$.  The simulation box is $60\lambda_0\times30\lambda_0$ and the gird size is $\Delta x=\triangle y=0.02\lambda_0$, and there are 200 macro-particles in each cell.

The main process involves the laser propagates forward along the axis ($+x$ direction) of the conical channel, due to the skin effect \cite{Diachenko:1979}, only when the laser intensity is strong enough, the electrons in the inner surface of wall are periodically extracted into the conical channel by the transverse laser field and accelerated forward by laser ponderomotive force.

Now let us demonstrate the simulation results of the two different cases. Fig.2(a) and 2(d) show the distribution of the electron density for the curved wall conical channel and for the flat one at $t=30T_0$. Some electrons are pulled from the wall by the laser transverse electric field, forming a sawtooth-like attosecond electron bunches with an interval of $\lambda_0$. The distribution of electrons on the upper and lower surfaces of the channel has a positional deviation of $\lambda_0/2$ due to the phase difference of $\pi$ between the upper and lower parts of the laser transverse electric field. It can be seen that the electron density in the curved wall conical channel reaches $22n_c$, which is significantly higher than the $8n_c$ observed in the flat channel. The laser electric field will be enhanced due to the focusing effect when the laser propagates in the conical channel. Figs.2(b) and 2(e) show that the transverse electric field in the curved wall conical channel is larger than that in the flat one. As a result, more electrons are pulled into the channel, achieving greater acceleration and resulting in a longer transverse length. Besides, the electron bunches in Fig.2(a) has a narrower longitudinal thickness as $\sim 300\rm{as}$ than in Fig.2(d). We attribute this difference to the longitudinal electric field as shown in Figs.2(c) and 2(f) for the curved wall conical channel and the flat one at $t=30T_0$, respectively. Note that the rectangular box in Fig. 2(a) corresponds to region A in Fig. 2(c), while the box in Fig. 2(d) corresponds to region B in Fig. 2(f). The electron bunch in region A experiences electric fields of comparable magnitude but opposite directions on both sides, whereas the bunch in region B encounters a larger negative electric field on the right side. This different longitudinal electric field behaviour results in the narrower longitudinal thickness of the electron bunches.

\begin{figure}[ht!]\centering
\includegraphics[width=0.32\textwidth]{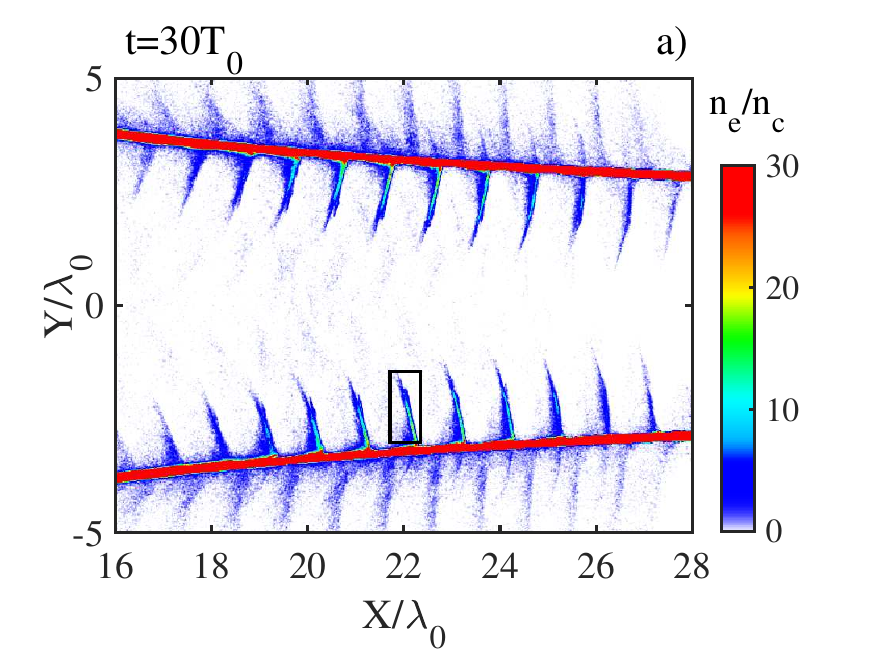}
\includegraphics[width=0.32\textwidth]{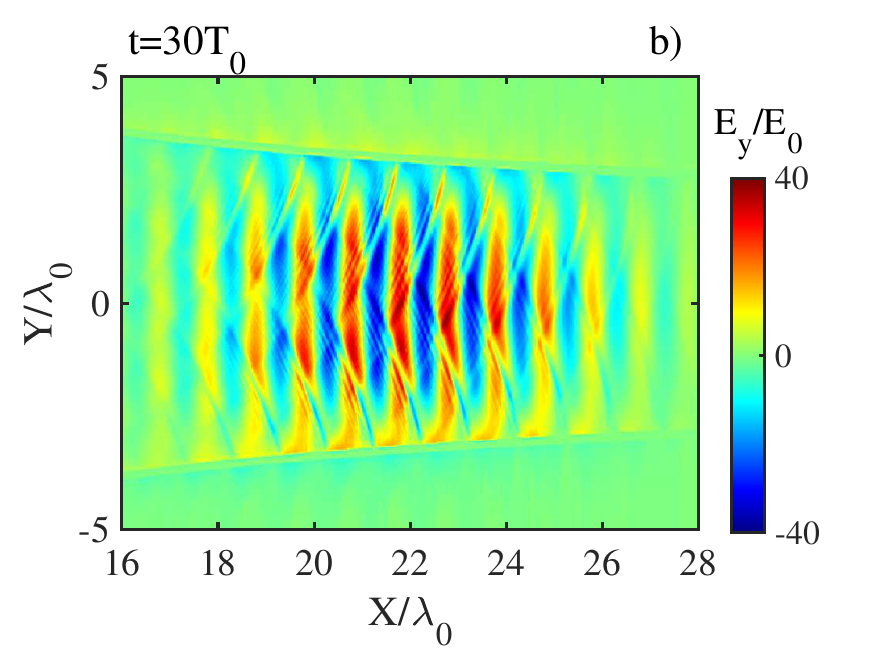}
\includegraphics[width=0.32\textwidth]{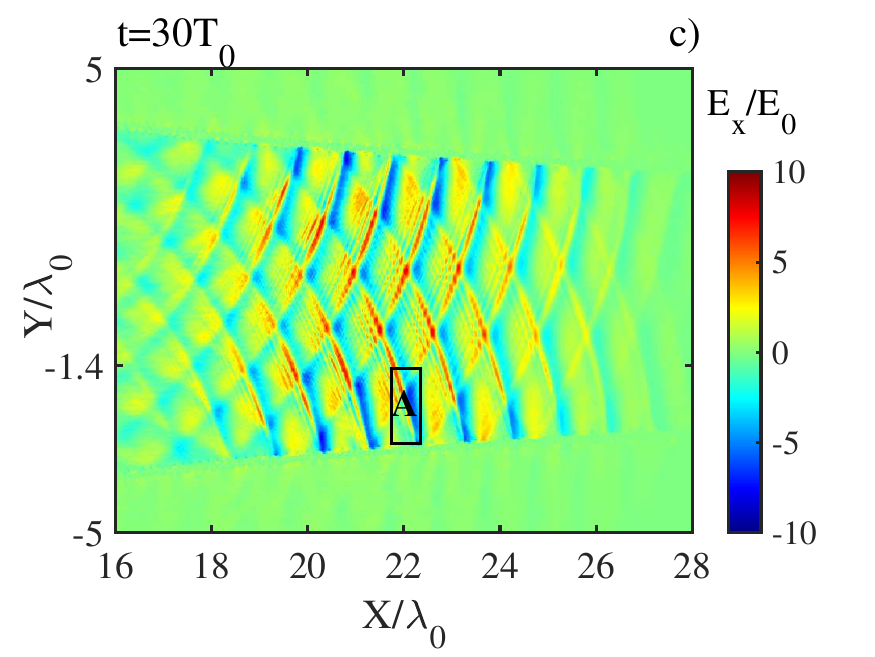}
\includegraphics[width=0.32\textwidth]{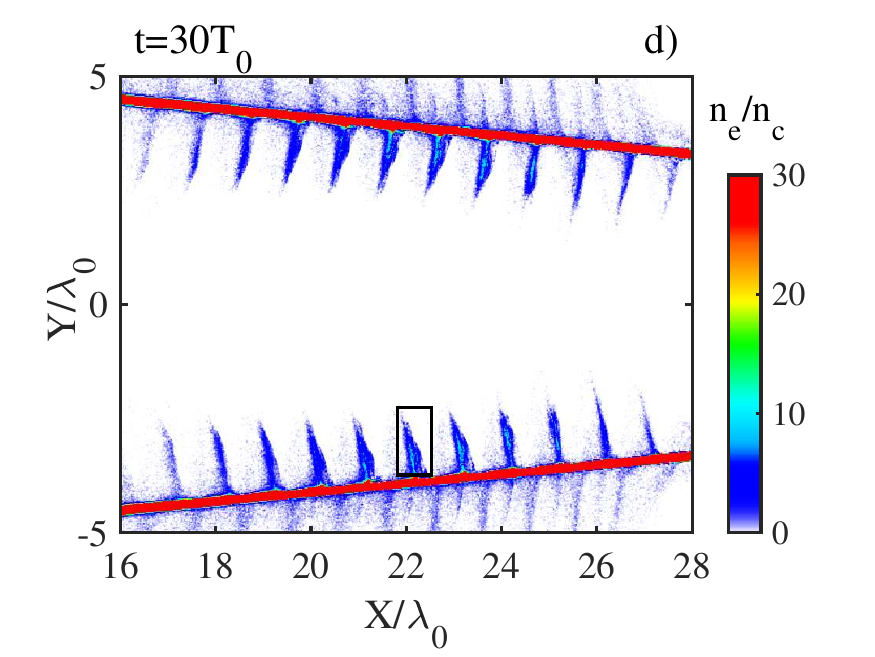}
\includegraphics[width=0.32\textwidth]{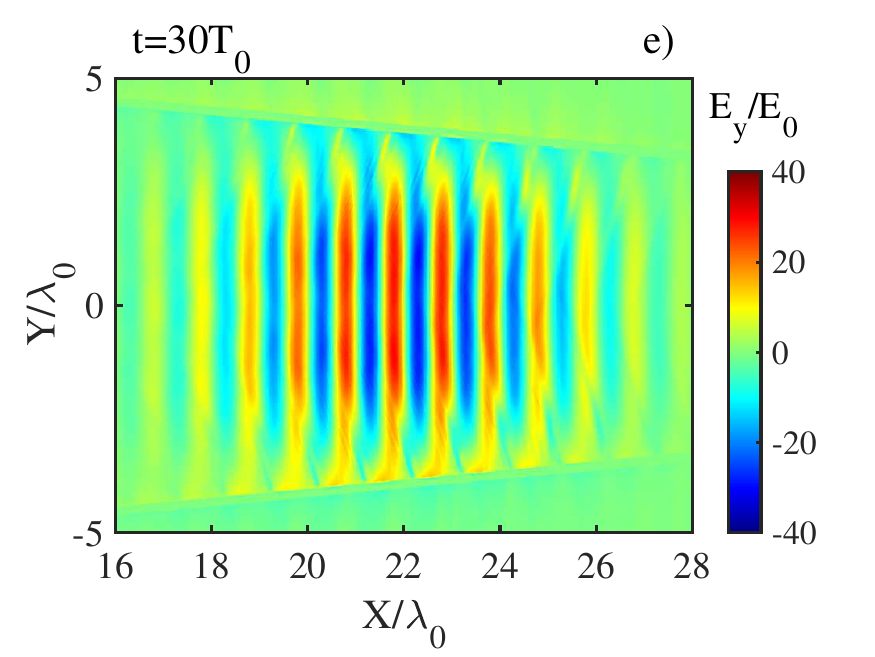}
\includegraphics[width=0.32\textwidth]{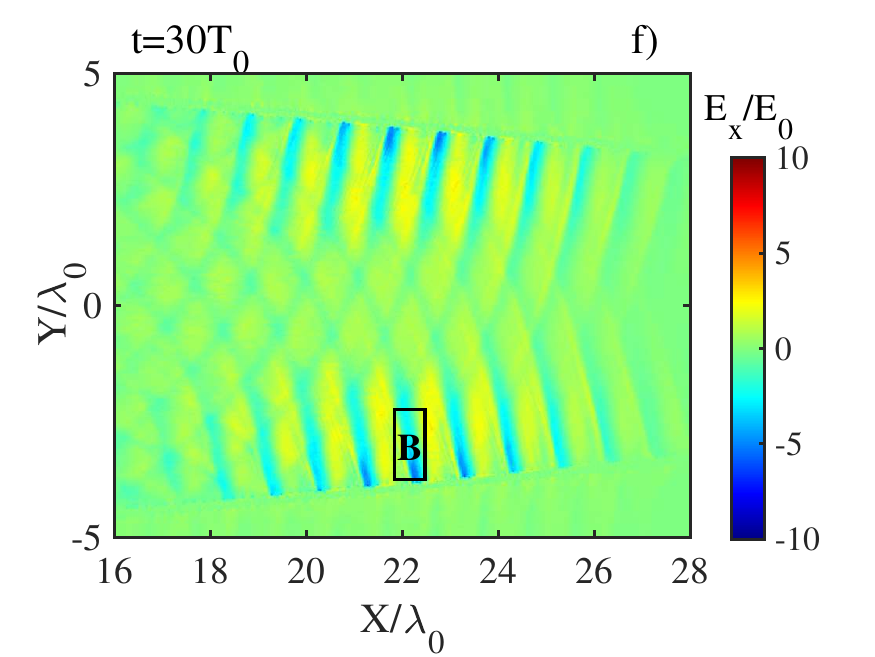}
\caption{The spatial distribution of electron density, transverse electric field and longitudinal electric field for the curved wall conical channel [(a),(b),(c)] and for the flat wall conical channel [(d),(e), (f)] in the $x-y$ plane at $t=30T_0$. Here, the electron density is normalized by $n_c$ and the electric field is normalized by $E_0$.
\label{fig:chap05fig6}}
\end{figure}

\begin{figure}[ht!]\centering
\includegraphics[width=0.49\textwidth]{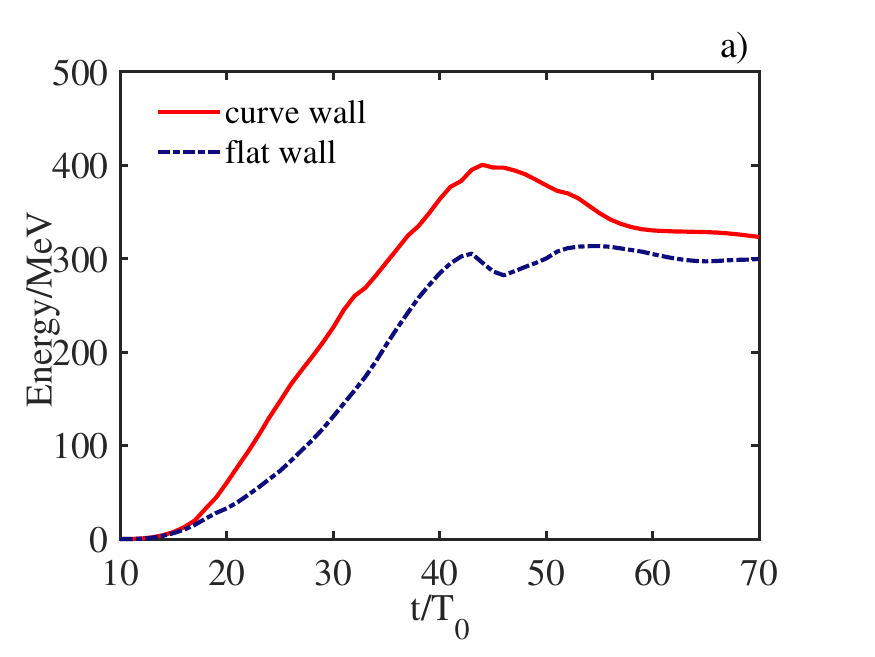}
\includegraphics[width=0.49\textwidth]{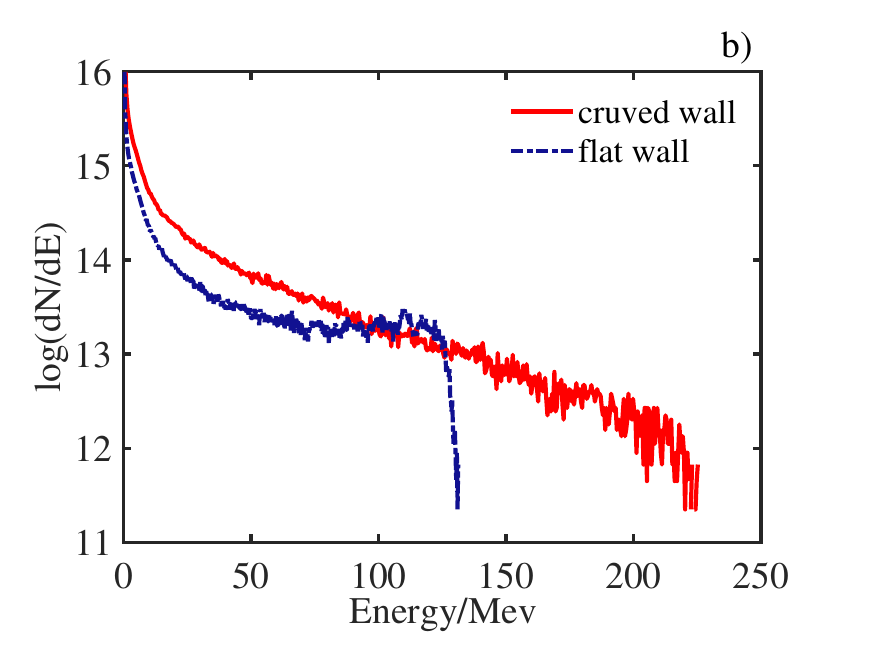}

\includegraphics[width=0.49\textwidth]{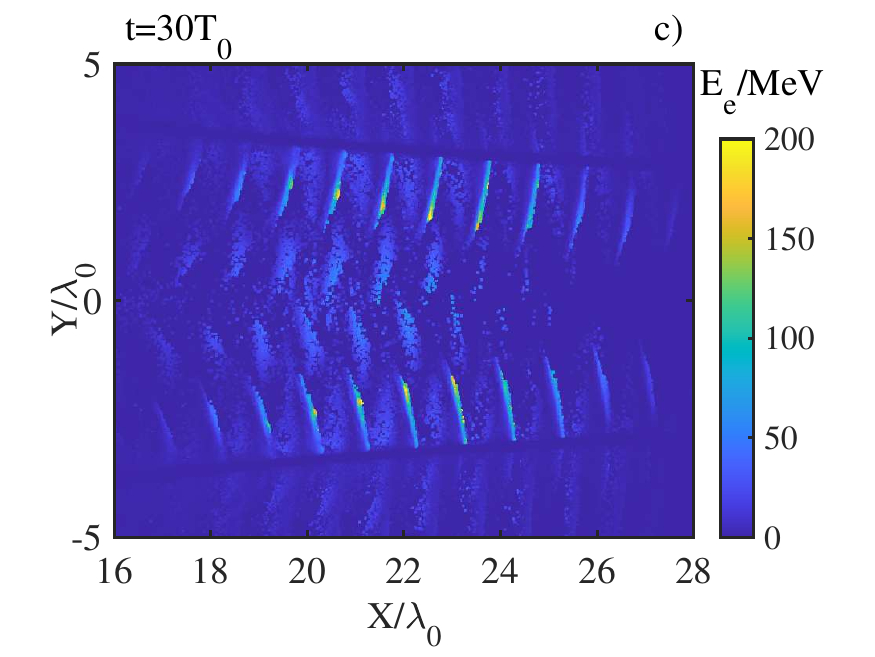}
\includegraphics[width=0.49\textwidth]{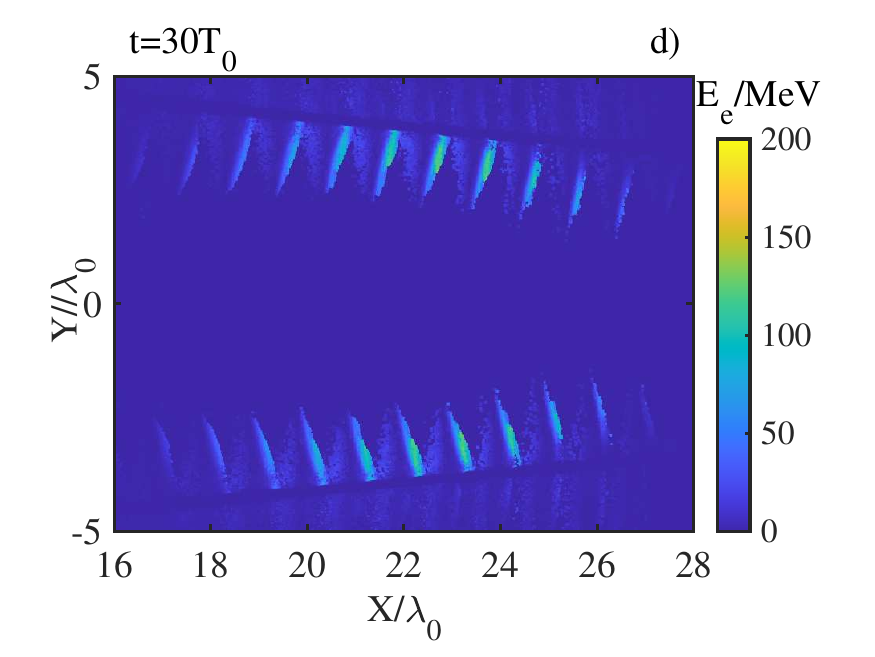}
\caption{The evolution of maximal electron energies (a) and the electron energy spectrum at $t=30T_0$ (b). The spatial distribution of electron energies in the $x-y$ plane at $t=30T_0$ for the the curved wall conical channel (c) and the flat wall conical channel (d).
\label{fig:chap05fig6}}
\end{figure}

To demonstrate the enhanced acceleration effect for the curved wall conical channel, the evolution of maximal electron energies is shown in Fig.3(a). The redline represents the electrons in the curved wall conical channel. As the laser propagates along the $+x$ direction, the intensity of the laser electric field increases, and electrons in both types of channels experience continuous acceleration when the peak of the laser electric field leaves the tip of the conical channel. For the electrons in the curved wall conical channel, their energy reaches maximal $\sim 400\rm{MeV}$ at $t=44T_0$, which is $\sim 100\rm{MeV}$ higher than that in the flat one of $\sim 300\rm{MeV}$ by an improving of $33 \%$.

Fig.3(b) shows the electron energy spectra at $t=30T_0$, where the redline is for the curved wall conical channel. The electron energy spectrum in the flat wall conical channel exhibits a flat distribution, with the number of electrons at the high-energy end being relatively uniform, and the maximum energy of electrons is about $130 \rm{MeV}$. However, the electron energy spectrum in the curved wall conical channel presents a typical exponential distribution, with the maximum energy $\sim 230\rm{MeV}$. This difference arises because the electrons in the curved wall conical channel experience a stronger local electric field, allowing more electrons are pulled into the channel and gain more energy directly from the electric field.

Figs.3(c) and 3(d) show the spatial distribution of electron energies in the $x-y$ plane at $t=30T_0$ for the curved wall conical channel and the flat one, respectively. It can be seen that the high energy electrons in both channels are distributed near the peak of the laser electric field, but the difference is that the high energy electrons in the curved wall conical channel are located at the tip of the bunches, and the high energy electrons in the other case are in the middle of the bunches. Once these electrons are pulled out in the channel, they are then accelerated forward by the laser ponderomotive force and obtain net energies from the laser. The electric field of the curved wall conical channel is stronger, so that the electrons being pulled out can get a greater acceleration, which results in the high energy electrons distributed at the tip of the electron bunches.

\begin{figure}[ht!]\centering
\includegraphics[width=0.24\textwidth]{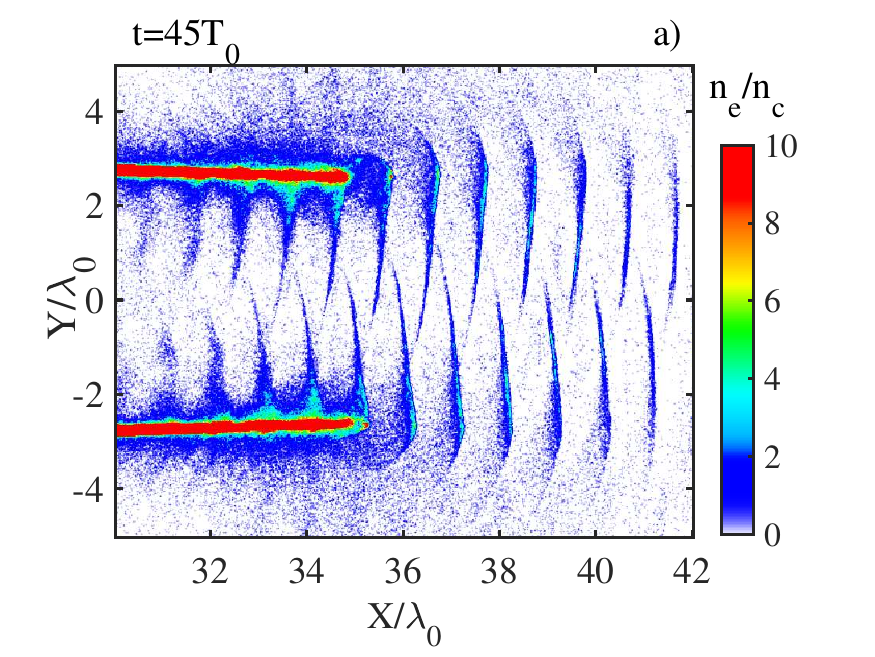}
\includegraphics[width=0.24\textwidth]{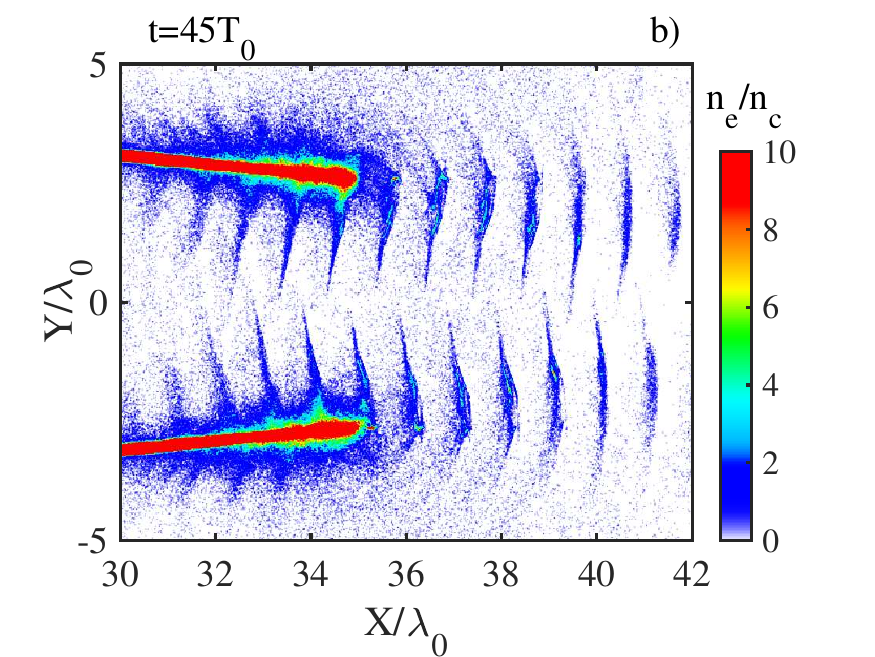}
\includegraphics[width=0.24\textwidth]{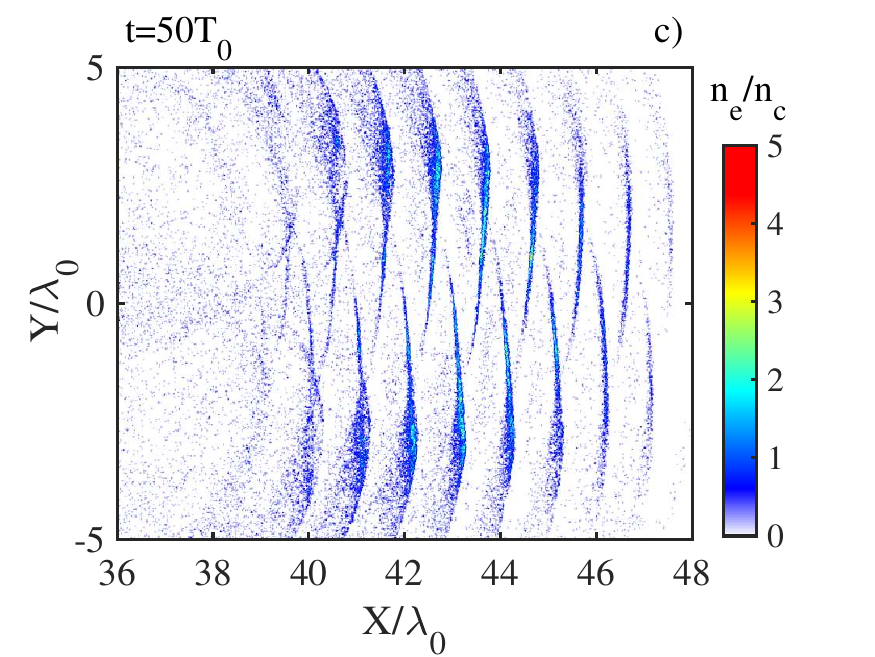}
\includegraphics[width=0.24\textwidth]{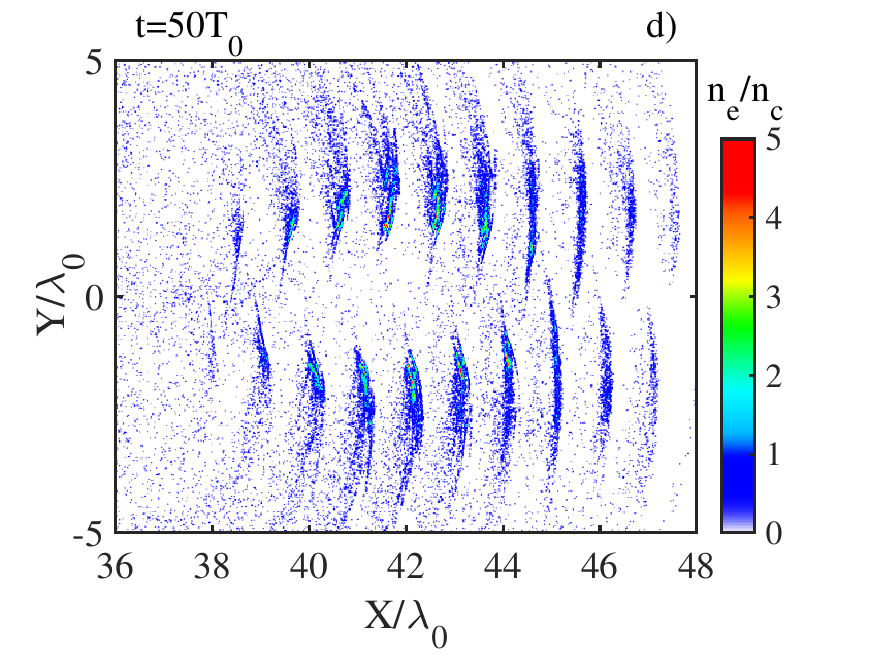}

\includegraphics[width=0.24\textwidth]{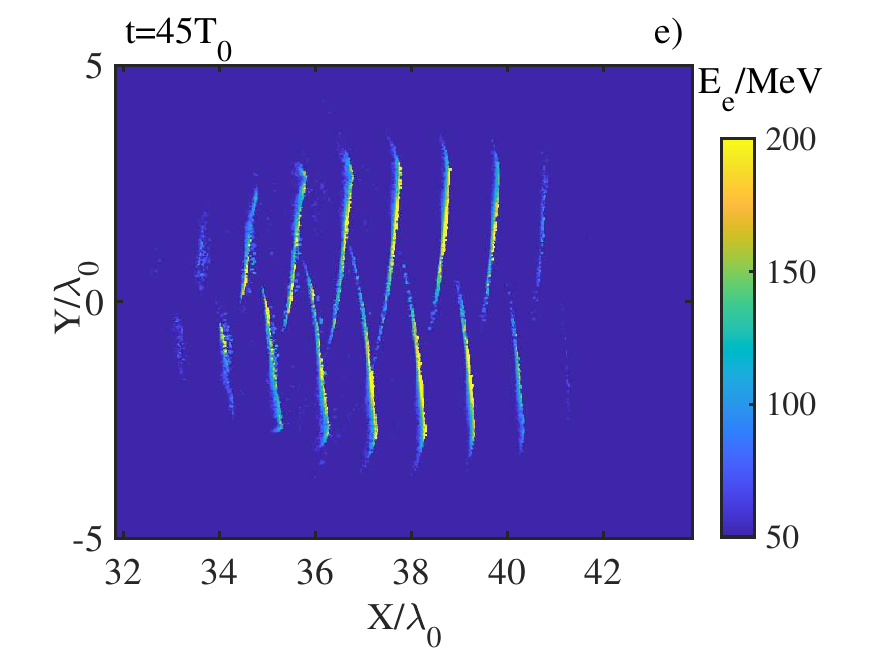}
\includegraphics[width=0.24\textwidth]{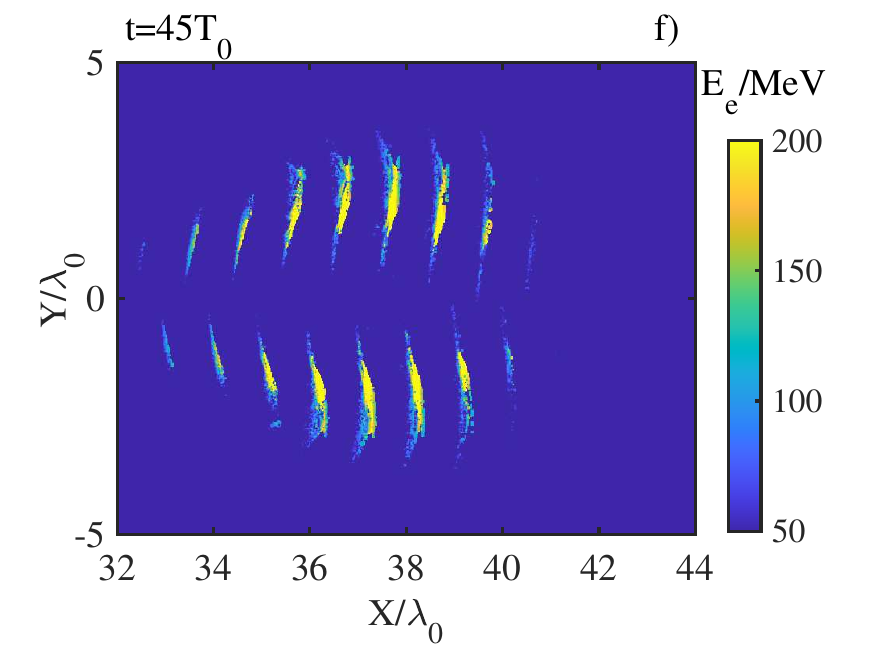}
\includegraphics[width=0.24\textwidth]{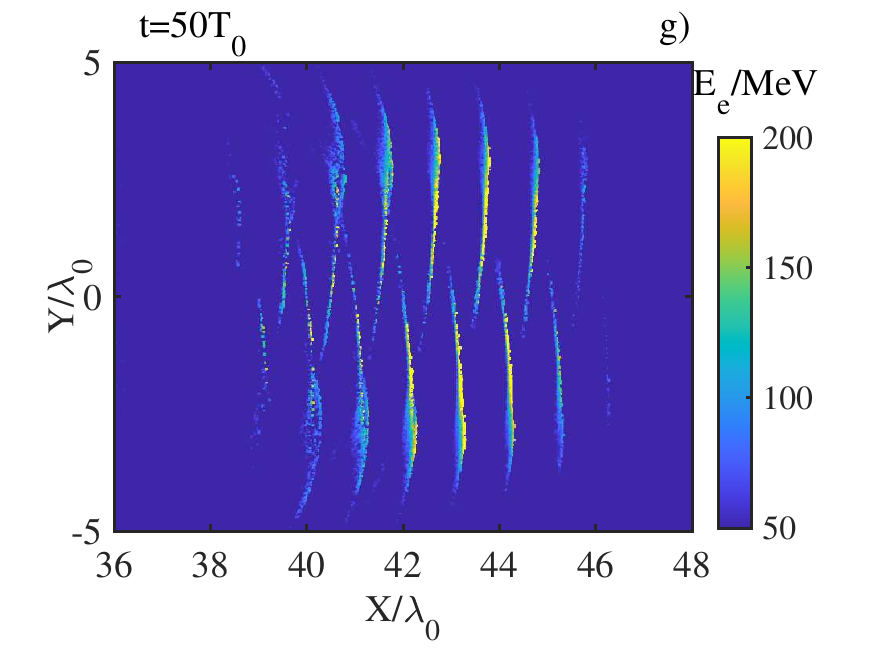}
\includegraphics[width=0.24\textwidth]{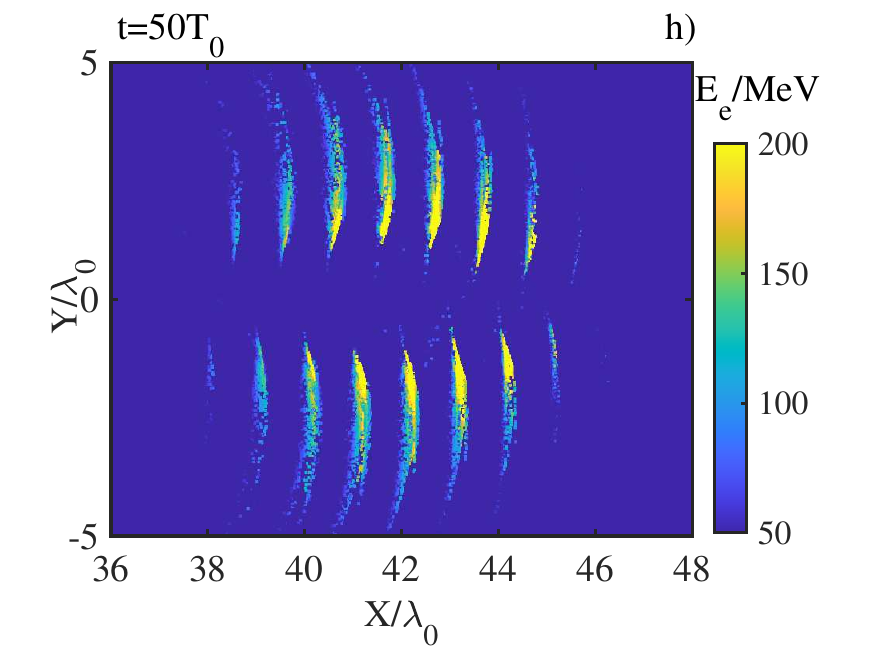}
\caption{The first row shows the spatial distribution of electron density and the second row displays the distribution of electron energy. The left two columns represent the distribution at $t=45T_0$ for the curve wall conical channel [(a), (e)] and the flat one [(b), (f)] and the right two columns represent the distribution at $t=50T_0$ for the curve wall conical channel [(c), (g)] and the flat one [(d), (h)].
\label{fig:chap05fig6}}
\end{figure}

In Fig.4, the first row shows the spatial distribution of electron density, while the second row displays the distribution of electron energy. The left two columns represent the distribution at $t=45T_0$ and the right two columns represent the distribution at $t=50T_0$. Figs.4(a), 4(c), 4(e) and 4(g) are the simulation results of the curved wall conical channel. The electrons continue to move forward while maintaining a sawtooth-like structure after leaving the channel. The longitudinal thickness of the electron bunch is about $200\rm{as}$ and $250\rm{as}$ at $t=45T_0$ and $t=50T_0$ for the curved wall conical channel, the corresponding electron bunches density is $4n_c$ and $2n_c$, respectively. We found that in both cases, the average energy of the electrons near the peak of the laser electric field is $170\rm{MeV}$. At $t=50T_0$, for the curved wall conical channel, the high energy electrons are mostly in the middle of the electron bunches, while in the flat wall conical channel, the high energy electrons are mostly in the tip position.

\begin{figure}[ht!]\centering
\includegraphics[width=0.49\textwidth]{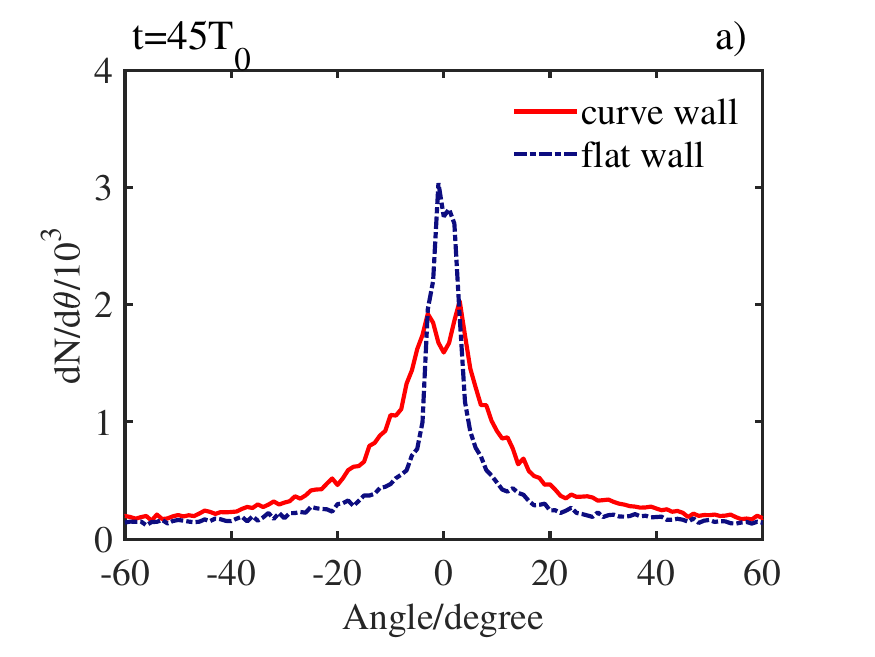}
\includegraphics[width=0.49\textwidth]{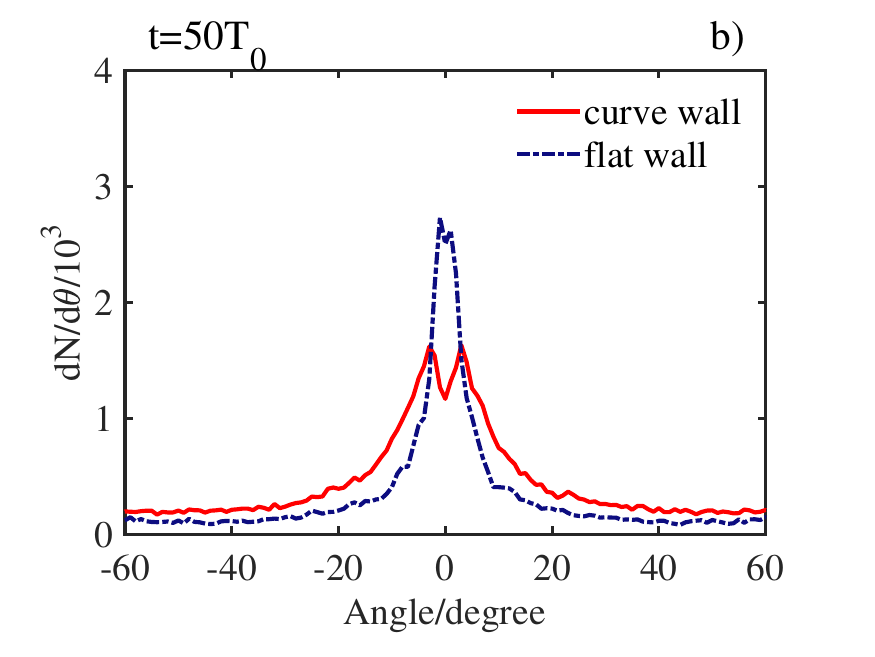}
\caption{The emission angle distribution of attosecond electrons emitted from the conical channel at $t=45T_0$ (a) and $t=50T_0$ (b).
\label{fig:chap05fig6}}
\end{figure}

To further investigate the behavior of the electron bunches after leaving the channel, Figs.5(a) and 5(b) show the emission angle distribution of attosecond electrons emitted from the conical channel at $t=45T_0$ and $t=50T_0$. The redline represents the electrons in the curved wall conical channel. The average emission angle
for the curved wall conical channel is about $9^{\circ}$ and for the flat one is about $4^{\circ}$. Furthermore, We also compare the brightness of six electron bunches in the two cases at $t=50T_0$, as shown in Table I. It can be seen that the brightness of the electron bunches in the first case is better, with a maximum difference of one order of magnitude for some cases. These electron bunches with high brightness have important applications in many fields such as medical imaging,  semiconductor manufacturing, nuclear fusion and so on.

\begin{table}[htbp]
 \setlength\tabcolsep{3pt}
    \caption{\label{tab1}The brightness of the attosecond electron bunches in the curved wall concical channel in the first column and the flat one in the second column at $t=50T_0$. }
    \begin{center}
        \begin{tabular*}{15cm}{@{\extracolsep{\fill}}l c c c c c c c c c c}
            \\ \hline\hline
            &curve wall & $2.74\times10^{27}$ & $4.78\times10^{27}$ & $1.14\times10^{28}$ & $2.40\times10^{27}$ & $8.53\times10^{27}$ & $1.10\times10^{28}$
               \\ \hline
            &plat wall & $1.38\times10^{27}$ & $8.23\times10^{26}$ & $3.35\times10^{27}$ & $1.08\times10^{27}$ & $1.29\times10^{27}$ & $8.36\times10^{26}$
            \\ \hline \hline
        \end{tabular*}
    \end{center}
\end{table}

\section{Other types of curved wall conical channels}

\begin{table}[htbp]
 \setlength\tabcolsep{3pt}
    \caption{\label{tab1}The density of electron at $t=30T_0$, the maximum energy of electron of three cases of circle, parabola and hyperbola.}
    \begin{center}
        \begin{tabular*}{15cm}{@{\extracolsep{\fill}}l c c c c c c c c c c}
            \hline\hline
            & case & density($ n_c$) & maximum energy($\text MeV$) 
            \\ \hline
            &circle & $25$ & $412$  &  $$\\ \hline
            &parabola & $22$ & $410$  & $$\\ \hline
            &hyperbola-II& $25$ & $378$  & $$
            \\ \hline \hline
        \end{tabular*}
    \end{center}
\end{table}

As shown above, the curved wall conical channel enhances the attosecond electron bunch generation and effectively improves the bunch quality. In this section, We have performed comparative simulation for several different curves as the walls of the conical channel while keeping other parameters the same. The geometric shape of the curve is a circle, a parabola, and a hyperbola with another form $y=a/(x+b)$, where $a$ and $b$ are constants, in order to distinguish between the two hyperbolic models, this form is called hyperbola-II.

Table II lists the electron density at $t=30T_0$ and the maximum energy of electron in evolution in several cases. The average of electron bunches density of the three cases exceeds $20n_c$, with the circular and hyperbolic-II cases achieving a density of 25$n_c$. Additionally, the maximum energy of the electron bunch is significantly enhanced compared to the flat wall conical channel, with increasing of $36\%$ observed for both the circular and parabolic cases.

Note that the variation in electron energy for the circular and parabolic curved walls are similar to the two cases of hyperbolic wall. After the laser peak leaves the channel, the maximum energy of the electrons decreases from the maximum to $370\rm{MeV}$ and $355\rm{MeV}$, respectively, within $10T_0$. However, the electron energy in hyperbolic-II keeps stable around the maximum electron energy of $380\rm{MeV}$, which would be beneficial for practical applications in experiments. We also found that in all three cases, high energy electrons are distributed near the peak of the laser electric field and at the tip of the bunches, which are consistent with previous studies.

\begin{figure}[ht!]\centering
\includegraphics[width=0.32\textwidth]{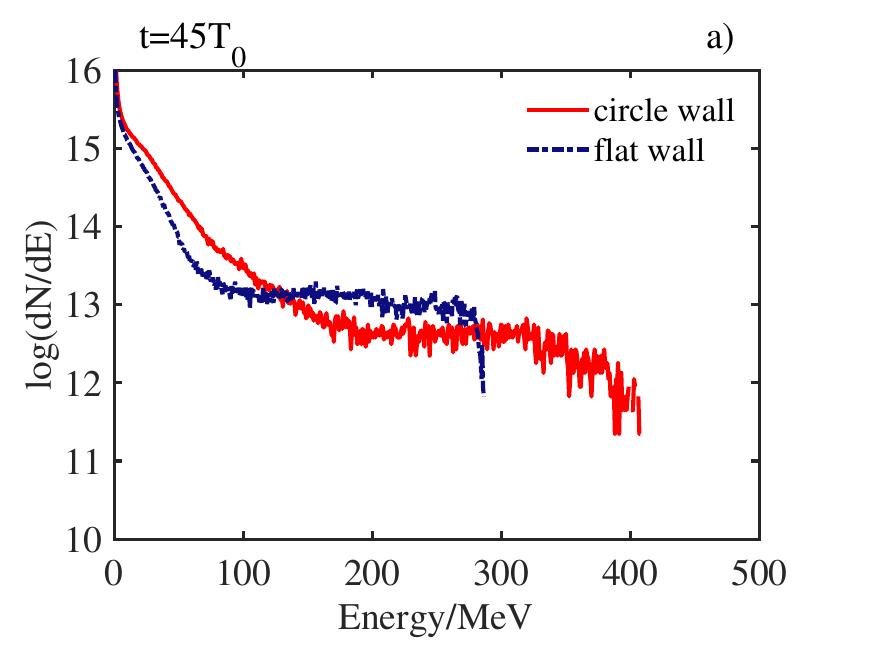}
\includegraphics[width=0.32\textwidth]{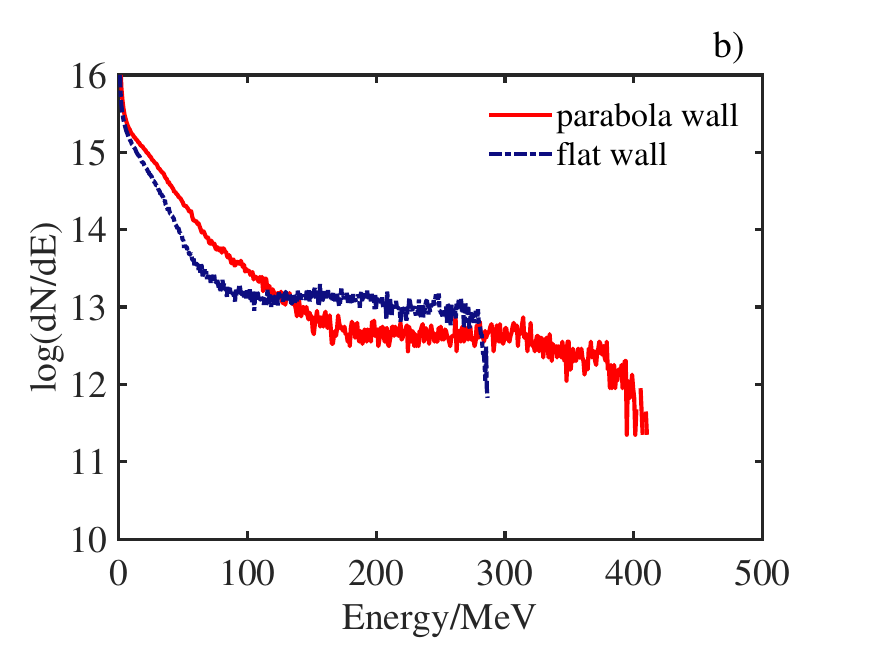}
\includegraphics[width=0.32\textwidth]{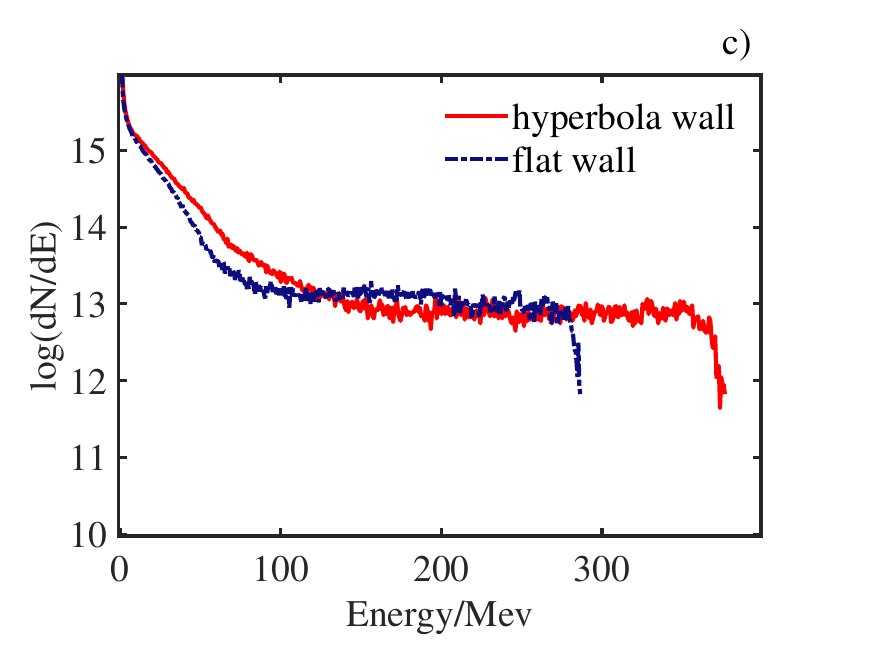}
\caption{Electron energy spectrum at $t=45T_0$ for the different geometric shapes of the curve circular (a), parabolic (b) and hyperbolic-II (c).
\label{fig:chap05fig6}}
\end{figure}

Fig.5 shows the electron energy spectrum distribution at $t=45T_0$ for the three cases. It can be seen that the electron energy spectrum generated in the circular and parabolic curve cases presents a typical exponential spectrum distribution, while the electron bunch generated in the hyperbolic-II curve case presents a flat distribution. The number of electrons with energies around $370\rm{MeV}$ is an order of magnitude higher compared to the other two shapes.

\section{Conclusion}

In conclusion, through the interaction of intense laser with the curved wall conical channel, we have acquied spatially periodic attosecond electron bunches with an average density of about $20 n_c$ and cut-off energy of $400\rm{MeV}$. Compared to the flat wall conical channel, the curved wall conical channel has better focusing effect while maintaining a small divergence angle. This leads to a $36\%$ increase in electron energy and a $175\%$ increase in average electron density. Such improvements are beneficial to the potential applications where a highly collimated attosecond electron bunch is required.

Furthermore, it is found that the geometric shape of the curved wall also affects the energy and density of the electron bunch. Among these, the electron energy obtained in the circular, parabolic and hyperbolic-I schemes has a higher value, but begins to drop sharply as the laser leaves the right opening of the channel. The electron energy obtained in the hyperbola-II still remains near the maximum value within $10 T_0$ after leaving the channel opening. The obtained attosecond electron hunch present in this study may have a wide range of applications in ultrafast optics, high energy physics, materials science and other fields in the future.

\begin{acknowledgments}
This work was supported by the National Natural Science Foundation of China (NSFC) under Grants No. 12375240 and No. 11935008. The computation was carried outat the HSCC of the Beijing Normal University.	
\end{acknowledgments}

\end{document}